\documentclass{revtex4}
\usepackage{t1enc}
\usepackage[latin1]{inputenc}
\usepackage{amsmath}
\usepackage{amssymb}
\usepackage{citesort}

\newcommand{\cM}{{\mathcal M}}

\newcommand{\del}{\partial}
\newcommand{\dal}{\square}

\newcommand{\eps}{\varepsilon}

\newcommand{\2}{\tfrac12}
\newcommand{\3}{\tfrac13}
\newcommand{\4}{\tfrac14}

\renewcommand{\i}{{\rm i\,}}

\begin{document}

\title{On the Velo-Zwanziger phenomenon}
\author{J\"org Frauendiener}
\address{Institut f\"ur Theoretische Astrophysik,
Universit\"at T\"ubingen,
Auf der Morgenstelle 10,
D-72076 T\"ubingen,
Germany}
\email{joergf@tat.physik.uni-tuebingen.de}
\begin{abstract}
The Rarita-Schwinger equation in a curved background and an
external electromagnetic field is discussed. We analyse the equation
in the 2-component spinor formalism and derive Buchdahl conditions for
them. The result is that the equation can consistently be imposed only
on Einstein manifolds with vanishing electromagnetic field.
\end{abstract}
\maketitle

\section{Introduction}
\label{sec:intro}

The Rarita-Schwinger equation~\cite{raritaschwinger41} has a number of
peculiar properties.  In the massless case the equation can be
regarded as one of Dirac's relativistic wave
equations~\cite{dirac36:_relat} for particles with spin $3/2$.
Fierz~\cite{fierz39:_ueber_theor_teilc_spin,fierz40:_ueber_drehim_teilc_ruhem_spin}
had pointed out that there exist hierarchies of such equations which
describe the same one-particle states. One can move within the
hierarchy by taking appropriate derivatives. In this sense the
Rarita-Schwinger field is related to the usual zero rest mass field
equation for spin $3/2$~\cite{penrose84:_spinor_spacet_i} by one
derivative. Fierz also observed that in general the solutions of these
equations are not unique but only defined up to `gauge solutions'
which do not contribute to the energy and angular-momentum expressions
constructed from the fields. This is the case also for the massless
Rarita-Schwinger field.

A solution of the massless Rarita-Schwinger equation gives rise to
such a `potential modulo gauge' description of a spin $3/2$ field in
close analogy to the Maxwell (i.e. the spin $1$)  case, where the
electromagnetic field can be obtained from a potential which itself
satisfies but is not completely determined by a field equation. There
is still a gauge freedom present which must be fixed before the
potential can be uniquely determined by its field equation.

While the (direct) formulation for a spin $3/2$ field in terms of the
conventional zero rest mass equation becomes inconsistent in any
conformally curved space-time (see
e.g.~\cite{penrose84:_spinor_spacet_i}) this is not so for the
`potential modulo gauge' description. Here again there are consistency
conditions to be satisfied but in this case they involve only the
Ricci tensor. This is a remarkable fact because the Rarita-Schwinger
equation seems to be the only system of spinor equations where
only the Ricci tensor appears in the obstruction to consistency. Thus,
the vacuum Einstein equations can be considered as being
`integrability conditions' for the massless Rarita-Schwinger
equation. This observation has been one motivation for an attempt to
reconcile twistor theory with arbitrary (vacuum)
space-times~\cite{penrose93:_twist_einst}. It also plays a fundamental
role in the theory of super-gravity.

The massive case has been studied by various authors, in particular,
by Velo and
Zwan\-zi\-ger~\cite{velozwanziger69:_propag_quant_rarit_schwin} who
discuss the massive Rarita-Schwinger field coupled to an external
electromagnetic field and by Madore~\cite{madore75:_einst_maxwel} who
in addition coupled the field to a linearized gravitational
field. Their result is that in these circumstances the
Rarita-Schwinger field seems to propagate acausally in the sense that
the characteristics of the equation become space-like so that
information about the field configuration can travel at speeds larger
than the speed of light. This effect has been termed the
`Velo-Zwanziger phenomenon'.

In this paper we offer a new analysis of the massive Rarita-Schwinger
equations on a curved background in an exterior electromagnetic
field. The plan of the paper is as follows: in
sect.~\ref{sec:translation} we first translate the equations into the
2-component spinor formalism decomposing fields and equations into
irreducible parts. Then, in sect.~\ref{sec:31} we derive the $3+1$
decomposition of these equations in the special case of flat Minkowski
space-time and vanishing electromagnetic field using the space-spinor
formalism. We show that the fields are necessarily undetermined. While
in the massless case the indeterminacy can be circumvented this is no
longer true in the massive case. Furthermore, in
sect.~\ref{sec:buchdahl} we derive Buchdahl conditions for the
equations which show that consistency of the equations holds only in
Einstein spaces with no electromagnetic field. This is in contrast to
the earlier results because it implies that there is no Velo-Zwanziger
phenomenon because one cannot couple the field consistently to a
electromagnetic field in the first place.

\section{Translation to 2-component spinors}
\label{sec:translation}

The Rarita-Schwinger equation~\cite{raritaschwinger41} in an external
electromagnetic field was formulated
in~\cite{velozwanziger69:_propag_quant_rarit_schwin} in terms of Dirac
spinors:
\begin{equation}
  \label{eq:vzorig}
  \left(\i \Gamma\cdot\nabla - B \right)_a{}^d \psi_d = 0.
\end{equation}
Here, $\psi_d$ is a 1-form taking values in a `charged bundle' of Dirac
spinors over space-time $\cM$. This is the usual Dirac bundle equipped
with an action of the group $U(1)$. Correspondingly, the space-time
connection, denoted by $D_a$, is promoted to a `charged connection'
$\nabla_a$ by `minimal coupling', $\nabla_a = D_a - \i
eA_a$. In~\eqref{eq:vzorig} the differential
operator is given by the term
\begin{equation}
  \label{eq:Gammapi}
\i\Gamma \cdot \nabla = \i\gamma^5\eps_{ab}{}^{cd}\nabla_c\gamma^b.
\end{equation}
In order to derive and discuss the consistency conditions for this
equation it is useful to translate it first into the formalism of
2-component spinors as presented
in~\cite{penrose84:_spinor_spacet_i}. Note that we use these
conventions throughout the paper.

We represent the Dirac spinor valued 1-form $\psi_d$ as
\begin{equation}
  \label{eq:dirac1form}
  \psi_d = \left( 
    \begin{array}{c}
      \phi_{DD'S}\\
      \chi_{DD'S'}
    \end{array}
    \right).
\end{equation}
The Dirac matrices are represented in the form given
in~\cite{penrose84:_spinor_spacet_i} as\footnote{Note that the
Clifford relation obeyed by the Dirac matrices used in
\cite{velozwanziger69:_propag_quant_rarit_schwin} and in
\cite{penrose84:_spinor_spacet_i} differ by a sign. This is
compensated for in the formulae}
\begin{equation}
  \label{eq:gammas}
  \gamma_a=\sqrt2 \left( 
    \begin{array}{cc}
      0 & \eps_{AR}\eps_{A'}{}^{S'}\\
      \eps_{A'R'}\eps_A{}^S & 0
    \end{array}
    \right),\hspace{1cm}
    \gamma^5 = \left( 
    \begin{array}{cc}
      -\i \eps_R{}^S & 0\\
      0 & \i\eps_{R'}{}^{S'}
    \end{array}
    \right).
\end{equation}
The mass term in (\ref{eq:vzorig}) is given by $B =
B_a{}^d = m \gamma_a{}^b$ where $\gamma_{ab}$ are the matrices
\begin{equation}
  \gamma_{ab} = \2\left( \gamma_a \gamma_b - \gamma_b \gamma_a\right).
\end{equation}
Represented in terms of 2-component spinors these matrices read
\begin{equation}
  \label{eq:gamma2}
  \gamma_{ab} = 2 
  \left(
    \begin{array}{cc}
      \eps_{A'B'} \eps_{R(A} \eps_{B)}{}^S & 0\\
      0& \eps_{AB} \eps_{R'(A'} \eps_{B')}{}^{S'}
    \end{array}
  \right).
\end{equation}
Inserting these representations into~(\ref{eq:vzorig}) we obtain after
some calculation the system of equations
\begin{align*}
  \nabla_{AB'}\chi_{BA'}{}^{B'} -   \nabla_{BA'}\chi_{AB'}{}^{B'}  &=
  m \phi_{A'AB},\\
  \nabla_{BA'}\phi_{B'A}{}^{B} -   \nabla_{AB'}\phi_{A'B}{}^{B}  &=
  m \chi_{AA'B'}.
\end{align*}
Note, that in the case $m=0$ the two equations decouple. If we assume
for the moment that the space-time is Minkowski space and that there
is no electromagnetic field present then it is obvious that there is
an arbitrariness in these equations. We are free to replace
$\chi_{AA'B'}$ and $\phi_{AA'B}$ with $\chi_{AA'B'} + \nabla_{AA'}
\chi_{B'}$ and $\phi_{AA'B} + \nabla_{AA'} \phi_{B}$ for arbitrary
spinor fields $\chi_{B'}$ and $\phi_B$ without changing the
equations. Thus, whenever $\chi_{AA'B'}$ and $\phi_{AA'B}$ are
solutions then so are $\chi_{AA'B'} + \nabla_{AA'} \chi_{B'}$ and
$\phi_{AA'B} + \nabla_{AA'} \phi_{B}$. This is a well-known property
of the massless Rarita-Schwinger
equations~\cite{penrose93:_twist_einst,dirac36:_relat,fierz39:_ueber_theor_teilc_spin,fierzpauli39}. It
implies that a solution of the equations for $m=0$ can be determined
only \emph{up to gauge transformations} of the above form and the
equations can be regarded as determining a spin $3/2$ field in a
`potential modulo gauge' description.

In the case $m\ne0$ the same replacement does not yield a new solution
because now the equations imply that $\chi_{B'}$ has to be covariantly
constant. So in this case there is no notion of a field being given by
a potential modulo gauge description. 

The final step in rewriting the Rarita-Schwinger equation is to
decompose the fields and the equations into irreducible parts. Thus,
we write  
\begin{align*}
  \phi_{AA'B} = \sigma_{A'AB} + \eps_{AB}\,\sigma_{A'}\\
  \chi_{AA'B'} = \tau_{AA'B'} + \eps_{A'B'}\,\tau_{A}
\end{align*}
where now the fields $\sigma_{AA'B}$ and $\tau_{AA'B'}$ are symmetric
in their last pair of indices. Then we obtain the following system of
spinor equations
\begin{align}
  \nabla_{B'(A}\tau_{B)A'}{}^{B'} - \nabla_{A'(A}\tau_{B)} = m\,
  \sigma_{A'AB},\label{eq:dtau3}\\
  \nabla_{BB'}\tau^{BB'}{}_{A'} + 3 \nabla_{BA'}\tau^B = -6\, m\,
  \sigma_{A'},\label{eq:dtau} \\
  \nabla_{B(A'}\sigma_{B')A}{}^{B} - \nabla_{A(A'}\sigma_{B')} = m\,
  \tau_{AA'B'},\label{eq:dsigma3}\\ 
  \nabla_{BB'}\sigma^{B'B}{}_{A} + 3 \nabla_{AB'}\sigma^{B'} = -6\, m\,
  \tau_{A}.\label{eq:dsigma}  
\end{align}
This is the set of equations we will analyse in the following
sections.

\section{$3+1$ decomposition}
\label{sec:31}

In order to find the basic propagation properties of this system of
equations we need to perform a $3+1$-splitting of the system. This is
done as usual using the space-spinor
formalism~\cite{sommers80:_space}. To this end we fix a timelike
vectorfield $t^{AA'}$, normalised by $t_at^a=2$, so that
$t_{AA'}t^{BA'}=\eps_A{}^B$. We use this vectorfield to `convert the
primed indices to unprimed ones'. Thus, for instance, we write the
field $\tau_{AA'B'}$ as
\begin{equation}
  \label{eq:decomptau}
  t^{A'}_Ct^{B'}_B\tau_{AA'B'} = t_{ABC} + 2\eps_{A(C} t_{B)}
\end{equation}
where $t_{ABC}$ is totally symmetric in all its indices. Similarly,
the field $\sigma_{A'AB}$ yields two irreducible parts $s_{ABC}$ and
$s_{A}$ and we set $\sigma_A=\sigma_{A'}t^{A'}{}_A$. The derivative operator
$\nabla_{AA'}$ can be written in the form
\[
\nabla_{AA'} = t_{AA'} \del - t_{A'}^C \del_{AC} \iff 
t^{A'}_B \nabla_{AA'} = \eps_{AB} \del + \del_{AB}.
\]
It is enough for our purposes to assume that the underlying space-time
is Minkowski space and that there is no electromagnetic field
present. Then we can arrange for the time-like vector field $t^a$ to
be covariantly constant. This has the consequence that all derivatives
of the vector field vanish and that the differential operators $\del$
and $\del_{AB}$ commute.

Inserting these decompositions into~(\ref{eq:dtau3}--\ref{eq:dsigma})
yields six equations which we group into four evolution equations
\begin{gather}
  \label{eq:evt3}
  \del t_{ABC} + \del_{(A}{}^D t_{BC)D} - \del_{(AB} \left(t_{C)}
    -\tau_{C)} \right) 
  = -m s_{ABC},\\ 
  \label{eq:evt1}
  \del \left(t_{A} + \tau_A \right) 
  - \3 \del_{AB}\left(t^B + \tau^B \right)  
  - \tfrac23 \del_{AB} \left(t^B - \tau^B \right) 
  = m \left(s_A - \sigma_A\right),\\ 
  \label{eq:evs3}
  \del s_{ABC} - \del_{(A}{}^D s_{BC)D} + \del_{(AB} \left(s_{C)} -
    \sigma_{C)} \right)
  = -m t_{ABC},\\ 
  \label{eq:evs1}
  \del \left(s_{A} + \sigma_A \right) 
  + \3 \del_{AB}\left(s^B + \sigma^B \right)  
  + \tfrac23 \del_{AB} \left(s^B - \sigma^B \right) 
  = m \left(t_A - \tau_A\right)
\end{gather}
and two constraint equations
\begin{gather}
  \label{eq:ct3}
  0 = T_A \equiv \del^{BC} t_{BCA} + 2 \del_{AB} \left(t^B + \tau^B
  \right) + 3 m  \left(\sigma_A + s_A\right). \\
  \label{eq:cs3}
  0 = S_A \equiv \del^{BC} s_{BCA} + 2 \del_{AB} \left(s^B + \sigma^B
  \right) - 3 m \left(t_A + \tau_A\right).
\end{gather}
Now we can see the peculiar behaviour of this system:
\begin{itemize}
\item We only get propagation equations for the sums $t^A+\tau^A$ and
  $s^A + \sigma^A$ while the differences $t^A-\tau^A$ and
  $s^A-\sigma^A$ do not evolve. Since they are not present in the
  constraints either we conclude that they are not determined by the
  system.
\item The evolution equations are symmetric hyperbolic which is easily
  verified. The characteristics for the first equation are the light
  cone and a timelike cone in the interior of the light cone, which is
  also the characteristic for the second equation. Therefore, for any
  values of the difference fields the Cauchy problem for the evolution
  equations is well-posed. Note, that if we would impose the condition
  that the difference fields should be linearly dependent on the sum
  of the fields then we can change the characteristics of the system
  in an almost arbitrary way. However, this should be taken as an
  additional hint that the difference of the fields is in a sense an
  unphysical feature of the equations.
\item In the case $m=0$ the equations decouple into two sets
  consisting of two evolution equations and one constraint each. Let
  us consider the three equations \eqref{eq:evt3},\eqref{eq:evt1} and
  \eqref{eq:ct3}. Here again, the difference $t^A-\tau^A$ is not
  determined by the equations. The gauge transformation mentioned in
  sect.~\ref{sec:translation} now translate into the transformations
  \[
  t_{ABC} \mapsto t_{ABC} + \del_{(AB} \chi_{C)},\quad t_A \mapsto
  t_A-\tfrac32 \del\chi_A + \2 \del_{AB}\chi^B,\quad \tau_A \mapsto -
  \tau_A - \del \chi_A + \del_{AC} \chi^C.
  \]
  One can use this gauge transformation to make $\tau_A=0$ (so that
  the original field $\chi_{AA'B'}$ is symmetric in its primed
  indices).  This yields a system of two evolution equations and one
  constraint equation which has a well-posed Cauchy problem (see
  e.g.~\cite{frauendiener95:_32_ricci}). In this way the
  indeterminacy in the system can be circumvented and one can still
  make sense of the equations. This argument can be generalized to
  Ricci flat space-times.
\item Taking a time derivative of $T_A$ and $S_A$, commuting
  derivatives and using the evolution equations yields after some
  calculation the equations
  \begin{gather}
    \del T_A - \3 \del_{AB}T^B + m S_A = -6m^2 \tau_A,\\
    \del S_A + \3 \del_{AB}S^B + m T_A = -6m^2 \sigma_A.
  \end{gather}
  This shows that in the case $m\ne0$ the constraints do not propagate
  unless the fields $\tau_A$ and $\sigma_A$ vanish. Hence, the fields
  $\chi_{AA'B'}$ and $\phi_{AA'B}$ have to be restricted by the
  algebraic condition of symmetry in the primed resp. unprimed indices
  which is formulated in terms of the Dirac spinor valued 1-form
  $\psi_d$ as $\gamma^a\psi_a=0$.
\end{itemize}

\section{Consistency conditions}
\label{sec:buchdahl}

The algebraic condition emerging in the previous section and the fact
that there are consistency conditions in the massless case indicate
that there are likely to exist such conditions also in this case.  Our
aim in this section is to derive these conditions for the fields which
occur when we try to impose the system of equations upon them on an
arbitrarily curved manifold and for a given exterior electromagnetic
field. To this end we take further covariant derivatives of the
equations, commute them in order to introduce curvature terms and try
to eliminate all second order derivatives of the fields. If this is
possible we will end up with an algebraic relation between the fields,
the curvature and the exterior electromagnetic field.

For later convenience we introduce the notation
\begin{equation}
  \label{eq:notation}
  \left[ \nabla_{AA'},\nabla_{BB'} \right] = \eps_{AB}
    \dal_{A'B'} + \eps_{A'B'} \dal_{AB},
\end{equation}
where, in contrast to~\cite{penrose84:_spinor_spacet_i}, the curvature
derivations $\dal_{AB}$ and $\dal_{A'B'}$ here also contain the
electromagnetic field $F_{AB}$.

We first take a derivative of~(\ref{eq:dtau3}).
\begin{multline*}
  m \nabla_C{}^{A'} \sigma_{A'BA} = \nabla_C{}^{A'} \nabla_{B'(A}
  \tau_{B)A'}{}^{B'} - \nabla_C{}^{A'} \nabla_{A'(A}\tau_{B)} \\
  = -\eps_{C(A} \dal^{A'B'}\tau_{B)A'B'} + \nabla_{B'(A}
  \nabla_C{}^{A'} \tau_{B)A'}{}^{B'} + \nabla_{CA'}
  \nabla^{A'}{}_{(A}\tau_{B)} \\
  = -\eps_{C(A} \dal^{A'B'}\tau_{B)A'B'} + \2 \nabla_{AB'}
  \left[ \nabla_C{}^{A'} \tau_{BA'}{}^{B'} \right] + \2 \nabla_{BB'}
  \left[ \nabla_C{}^{A'} \tau_{AA'}{}^{B'} \right]
  + \2 \eps_{C(A} \dal \tau_{B)} + \dal_{C(B} \tau_{A)}
\end{multline*}
Now we use~(\ref{eq:dtau3}) and (\ref{eq:dtau}) to obtain
\begin{multline*}
  m \nabla_C{}^{A'} \sigma_{A'BA} = -\eps_{C(A} \dal^{A'B'}\tau_{B)A'B'}  
  + \2 \eps_{C(A} \dal \tau_{B)} + \dal_{C(B} \tau_{A)}\\
  + \2 \nabla_{AB'} \left[ - \nabla^{B'}{}_{(C} \tau_{B)} - m
    \sigma^{B'}{}_{BC} - 3 \eps_{BC} \left( m \sigma^{B'} +
      \nabla_D{}^{B'} \tau^D\right) \right]\\
  + \2 \nabla_{BB'} \left[ - \nabla^{B'}{}_{(C} \tau_{A)} - m
    \sigma^{B'}{}_{AC} - 3 \eps_{AC} \left( m \sigma^{B'} +
      \nabla_D{}^{B'} \tau^D\right) \right]\\
 =  -\eps_{C(A} \dal^{A'B'}\tau_{B)A'B'}  + \2 \eps_{C(A} \dal
 \tau_{B)} + \dal_{C(B} \tau_{A)} \\
  - \4 \eps_{A(C} \dal \tau_{B)} - \2 \dal_{A(C}  \tau_{B)} 
  - \2 m \nabla_{AB'} \sigma^{B'}{}_{BC} - \tfrac32
  \eps_{BC} \left( m \nabla_{AB'} \sigma^{B'} + \nabla_{B'A}
    \nabla^{B'}{}_D \tau^D \right)\\
  - \4 \eps_{B(C} \dal \tau_{A)} - \2 \dal_{B(C}
  \tau_{A)} - \2 m \nabla_{BB'} \sigma^{B'}{}_{AC} - \tfrac32
  \eps_{AC} \left( m \nabla_{BB'} \sigma^{B'} + \nabla_{BB'}
    \nabla^{B'}{}_D \tau^D \right).
\end{multline*}
Collecting appropriate terms we get
\begin{multline*}
    m \nabla_C{}^{A'} \sigma_{A'BA} + m \nabla_{B'(B}
    \sigma^{B'}{}_{A)C} - 3 m \eps_{C(A} 
  \nabla_{B)B'} \sigma^{B'} = \\
  -\eps_{C(A} \dal^{A'B'}\tau_{B)A'B'} + \tfrac34 \eps_{C(A} \dal \tau_{B)} - \2
  \dal_{AB} \tau_C + \2 \dal_{C(A} \tau_{B)}
   + \tfrac32 \eps_{C(A} \nabla_{B)B'}  \nabla^{B'}{}_D \tau^D \\
   = -\eps_{C(A} \dal^{A'B'}\tau_{B)A'B'} + \tfrac34 \eps_{C(A}
     \dal \tau_{B)} - \2 \dal_{AB} \tau_C - \tfrac34 \eps_{C(A} \dal
     \tau_{B)} + \tfrac32 \eps_{C(A} \dal_{B)D} \tau^D\\
   = -\eps_{C(A} \dal^{A'B'}\tau_{B)A'B'} 
     - 2 \dal_{AB} \tau_C + 2 \dal_{C(A} \tau_{B)}.
\end{multline*}
Rewriting the left hand side yields
\begin{multline*}
  - m \eps_{C(B} \left(  \nabla^{B'D} \sigma_{B'A)D} + 3
    \nabla_{A)B'} \sigma^{B'} \right)
  = - \eps_{C(A} \dal^{A'B'} \tau_{B)A'B'} - 2 \dal_{AB} \tau_C + 2
  \dal_{C(A} \tau_{B)}. 
\end{multline*}
Using~(\ref{eq:dsigma}) we get
\begin{equation}
   6m^2 \eps_{C(B} \tau_{A)}
  = - \eps_{C(A} \dal^{A'B'} \tau_{B)A'B'} - 2 \dal_{AB} \tau_C 
  + 2 \dal_{C(A} \tau_{B)},
\end{equation}
which is equivalent to its contraction
\begin{equation}
  \label{eq:relation_contracted}
   6m^2\tau_{A}
  = - \dal^{A'B'} \tau_{AA'B'}  + 2 \dal_{AB} \tau^{B},
\end{equation}
We could also have taken a derivative of~(\ref{eq:dtau}) and gone
through the same procedure. Then we would have ended up with exactly
the same relation as above. Introducing now the explicit form of the
curvature derivations we obtain the consistency conditions
\begin{equation}
  \label{eq:conscond}
  6m^2\tau_A = \Phi_A{}^{BA'B'}\tau_{BA'B'} - 6 \Lambda \tau_A + \i e
  F^{A'B'} \tau_{BA'B'} - 2\i e F_{AB} \tau^B. 
\end{equation}
This relation couples the values of the fields $\tau_A$ and
$\tau_{AA'B'}$ with the curvature and the extrinsic electromagnetic
field if there should be one. A similar relation holds for the fields
$\sigma_{A'}$ and $\sigma_{AA'B}$. It is not immediately obvious that
such relations should exist because this depends very much on the
detailed structure of the equation in question.

\section{Discussion}

Let us now discuss the relation~(\ref{eq:relation_contracted}) for
various special cases. Suppose we are in Minkowski
space and suppose that~(\ref{eq:dtau3}--\ref{eq:dsigma}) hold. Since
all the commutators vanish we necessarily recover the condition
\begin{equation}
  \label{eq:condmink}
     6m^2 \tau_B = 0.
\end{equation}
Thus, either we have $m^2=0$ or $\tau_B=0$. So if we insist on a
massive field then this field cannot have a component $\tau_B$ and the
spinor field $\chi_{BB'A'}$ must be symmetric in its last pair of
indices. On the other hand, in the massless case, we may admit the
part $\tau_B$ but it does not play a role. Instead it corresponds to
the gauge freedom discussed in sect.~\ref{sec:translation}.

Focussing now on a massive field in a general space-time with an
electromagnetic field we have $\tau_B=0$ and~\eqref{eq:conscond}
reduces to
\[
  0 = \Phi_A{}^{BA'B'}\tau_{BA'B'} + \i e F^{A'B'} \tau_{BA'B'} ,
\]
which has to hold for all values of the field $\tau_{BA'B'}$. This is
a severe algebraic restriction on the field and/or the electromagnetic
and curvature fields. Looking back at the evolution
equations~\eqref{eq:evt3},~\eqref{eq:evt1} we see that these are
symmetric hyperbolic evolution equations for the field $\tau_{BA'B'}$
which will be uniquely determined given data on an initial
hypersurface. These data are subject to the constraint~\eqref{eq:ct3}.
We have not checked the propagation of the constraints in the general
case but it is rather likely, that propagation of the constraints is
guaranteed by the consistency conditions derived above.

Thus, if we insist on having the full freedom in specification of
initial data then the consistency condition has to hold for arbitray
$\tau_{BA'B'}$ and this then implies that
\begin{enumerate}
\renewcommand{\labelenumi}{$(\roman{enumi})$}
\item the space-time is an Einstein manifold, $\Phi_{ABA'B'}$,
\item the electromagnetic field vanishes, $F_{AB}=0$.
\end{enumerate}
While the first consequence is familiar from the massless case, the
second consequence is new. It implies that it is impossible to
couple a Rarita-Schwinger field to an electromagnetic field in a
consistent way. Thus, the acausal propagation properties attributed to
that system are not really there, because the system cannot be set up
in a consistent way. This result is independent of the curvature of
the space-time i.e., even in a flat background one cannot have an
electromagnetic field present.

\section*{Acknowledgment}

I wish to thank D. Giulini for drawing my attention to the peculiar
behaviour of the Rarita Schwinger fields and the Velo-Zwanziger
phenomenon.


\begin{thebibliography}{10}

\bibitem{raritaschwinger41}
W. Rarita and J. Schwinger, Phys. Rev. {\bf 60},  61  (1941).

\bibitem{dirac36:_relat}
P.~A.~M. Dirac, Proc. Roy. Soc. London A {\bf 155},  447  (1936).

\bibitem{fierz39:_ueber_theor_teilc_spin}
M. Fierz, Helv. Phys. Acta {\bf 12},  3  (1939).

\bibitem{fierz40:_ueber_drehim_teilc_ruhem_spin}
M. Fierz, Helv. Phys. Acta {\bf 13},  45  (1940).

\bibitem{penrose84:_spinor_spacet_i}
R. Penrose and W. Rindler, {\em Spinors and Spacetime} (Cambridge University
  Press, Cambridge, 1984), Vol.~1.

\bibitem{penrose93:_twist_einst}
R. Penrose,  in {\em Proceedings of the Plymouth conference}, edited by S.
  Huggett (Marcel Dekker, New York, 1994).

\bibitem{velozwanziger69:_propag_quant_rarit_schwin}
G. Velo and D. Zwanziger, Phys.~Rev. {\bf 186},  1337  (1969).

\bibitem{madore75:_einst_maxwel}
J. Madore, Phys. Lett. B {\bf 55},  217  (1975).

\bibitem{fierzpauli39}
M. Fierz and W. Pauli, Proc. Roy. Soc. London A {\bf 173},  211  (1939).

\bibitem{sommers80:_space}
P. Sommers, J. Math. Phys. {\bf 21},  2567  (1980).

\bibitem{frauendiener95:_32_ricci}
J. Frauendiener, J. Math. Phys. {\bf 36},  3012  (1995).

\end{thebibliography}
\end{document}